\documentclass[aps,prd,twocolumn,superscriptaddress,nofootinbib]{revtex4-2}

\usepackage{amsmath}
\usepackage{amssymb}

\usepackage{color}

\usepackage[obeyFinal,colorinlistoftodos]{todonotes}

\usepackage{printlen}

\usepackage[markifdraft,dirty=*]{gitinfo2}

\usepackage{afterpage}

\usepackage{graphicx}
\graphicspath{{./img/}{../img/}{../../img/}{./paper/img/}{../paper/img/}{../../paper/img/}}
\setkeys{Gin}{draft=false}

\usepackage{amsmath,amsthm,amssymb}

\usepackage{macros}

\newcommand{\dd}{\ensuremath \mathrm{d}}
\usepackage{commath}

\usepackage[bookmarks]{hyperref}

\usepackage[capitalize,noabbrev]{cleveref}

\crefname{equation}{equation}{equations}

\date{15 March 2023}

\begin{document}

\title{Simulations of multi-field ultralight axion-like dark matter}

\author{Noah Glennon}
\email{nglennon@wildcats.unh.edu}

\author{Nathan Musoke}
\email{nathan.musoke@unh.edu}

\author{Chanda Prescod-Weinstein}
\email{chanda.prescod-weinstein@unh.edu}

\affiliation{Department of Physics and Astronomy\\ University of New Hampshire, Durham, New Hampshire 03824, USA}

\begin{abstract}
    As constraints on ultralight axion-like particles (ALPs) tighten, models with multiple species of ultralight ALP are of increasing interest.
We perform simulations of two-ALP models with particles in the currently supported range~\cite{Marsh:2013ywa} of plausible masses.
The code we modified, \texttt{UltraDark.jl}, not only allows for multiple species of ultralight ALP with different masses, but also different self-interactions and inter-field interactions.
This allows us to perform the first three-dimensional simulations of two-field ALPs with self-interactions and inter-field interactions.
Our simulations show that having multiple species and interactions introduces different phenomenological effects as compared to a single field, non-interacting scenarios.
In particular, we explore the dynamics of solitons.
Interacting multi-species ultralight dark matter has different equilibrium density profiles as compared to single-species and/or non-interacting ultralight ALPs.  As seen in earlier work~\cite{Glennon:2020dxs}, attractive interactions tend to contract the density profile while repulsive interactions spread out the density profile.
We also explore collisions between solitons comprised of distinct axion species. We observe a lack of interference patterns in such collisions, and that resulting densities depend on the relative masses of the ALPs and their interactions.

\end{abstract}

\maketitle


\section{Introduction}
\label{sec:introduction}

There is strong evidence that approximately 85\% of the matter in the universe is practically invisible, so far only detected through its gravitational effects on luminous matter~\cite{Planck:2015fie, BERTONE2005279, Bertone2018, Freese:2008cz, Chabanier:2019eai}.
The nature of this dark matter is a matter of open investigation.

One such proposal is the QCD axion, a particle introduced to resolve the strong CP problem through the Pecci--Quinn mechanism~\cite{Dine:1981rt,Drlica-Wagner:2022lbd}.
These scalar particles have a mass of $10^{-11} \;\mathrm{eV}$--$10^{0} \;\mathrm{eV}$ for the QCD to be dark matter.  The QCD axion can be generalised to a class of axion-like particles (ALPs).
Among these these is fuzzy dark matter (FDM), a form of dark matter whose constituent particles have a mass of $\sim 10^{-22} \;\mathrm{eV}$~\cite{Hu2000, Hui:2016ltb, Ringwald:2012hr}.
This means that its de Broglie wavelength is very large and the small scale structure of dark matter halos is different from that expected from more massive particles.

Recent work has argued that a lack of observed gravitational heating in ultra-faint dwarf galaxies constrains the dark matter particle mass to $m > 3 \times 10^{-19} \;\mathrm{eV}$~\cite{Dalal:2022rmp}.
This constraint relies on an assumption that Segue 1 and Segue 2 are representative of other galaxies, and that a single FDM species comprises a majority of the dark matter.
The inclusion of multiple species of ultralight particle significantly reduces the expected gravitational heating~\cite{Gosenca:2023yjc}.

Ultralight axions (ULAs) are an extension of FDM models that include interactions.
Some have considered the effects of self interactions on such ultralight scalar models~\cite{Glennon:2020dxs, Chavanis:2016dab, Glennon:2022huu, Fan:2016rda, Unal:2020jiy, Chakrabarti:2022owq}.  Although constraints in the single-field scenario predict the self-interaction strength be very small, it should not be ignored since the phase-space density of axions in these systems is extremely large~\cite{Desjacques2017, Glennon:2022huu}.

Another common assumption is that a particular dark matter model accounts for all --- or at least a significant fraction --- of the dark matter in the universe.
Thus far, most work on ALPs not only assumes that the ALP comprises a significant portion of the dark matter, the ALP itself is comprised of a single species.
However, the assumption of a single field should be considered a toy model.
The generic prediction from string theory is of an ``axiverse'' of ALPs~\cite{Arvanitaki:2009fg}.
In such a scenario, there would be a multiple ALPs with a hierarchy of masses; we use the term \emph{species} or \emph{field} to refer to these different ALPs.

Recent work has begun to explore this part of theory space.
In~\cite{Luu:2018afg}, the authors examine stable time-independent solitonic solutions of multi-field models and argue that the existence of axion fields with multiple masses is a plausible explanation for observed dark matter substructure.
The properties of such nested solitons have been further studied in~\cite{Eby:2020eas}, with the addition of self-interactions.
In~\cite{Guo:2020tla}, the authors studied closely related multi-field boson stars, solving numerically for equilibrium solutions.  In the recent paper~\cite{Gosenca:2023yjc}, the authors simulate multi-field ULDM halos and find that introducing more particle species smooths out outer halo profile.

Not only could each of these fields have self-interactions, there is the possibility of inter-field interactions.
Interactions between the multiple fields opens up a range of novel phenomenologies.
In~\cite{Eby:2015hsq}, the authors say that higher order repulsive self-interaction terms may stabilize solitons from collapsing into black holes when they have lower order attractive self-interactions~\cite{Khlebnikov:1999qy}.
In single field models with attractive self-interactions, there is a maximum mass a soliton can have without collapsing into a black hole or becoming an axinova~\cite{Chavanis:2016dab, Fox:2023aat}.  
Repulsive inter-field interactions may remove such instabilities in nested solitons with attractive self-interactions.
Interactions between axions in the early universe can also lead to transfers of energy between species~\cite{Cyncynates:2021xzw,Cyncynates:2022wlq}.
Interactions between axions and other scalar ultralight fields such as dilatons in the early can also affect dark matter abundances~\cite{Xu:2022yxr}.

There have been a number of recent papers on structure formation and soliton condensation in single field FDM models~\cite{Kirkpatrick:2020fwd,Kirkpatrick:2021wwz,Levkov2018,Du:2016zcv}.
Structure formation in the axiverse is likely distinct from what is described in recent papers on structure formation and soliton condensation in single field FDM models. These papers have examined the timescales required for the condensation of stable configurations from an incoherent FDM field.
This process would be significantly altered by the existence of multiple fields.
In the most extreme case, multiple fields with repulsive interactions may even be partitioned early in structure formation, leading to different dark matter species in different galaxies.
This would have an effect on, for example, rotation curves and strong gravitational lensing. 

Multi-component dark matter would have an appreciable affect on gravitational lensing observables.
Relative time delays of strongly lensed systems are used to measure $H_0$, the expansion rate of the universe~\cite{Refsdal:1964nw,Wong:2019kwg,DES:2019fny,Millon:2019slk}.
Recent work has explored the systematic uncertainties in these measurements due to a mass-sheet degeneracy~\cite{Blum:2020mgu,Yildirim:2021wdd,Blum:2021oxj,Birrer:2020tax}.
In particular, a $m \sim 10^{-25} \text{eV}$ particle comprising $\sim 10\%$ of the dark matter could have a significant effect on the inferred $H_0$, highlighting the relevance of multi-species models~\cite{Blum:2021oxj}. Recent work has explored the possibility of higher-spin ultralight bosonic dark matter and has shown that it can have similar phenomenology to ultralight ALPs~\cite{Jain:2021pnk,Amin:2022pzv,Jain:2022agt}.
In fact, it has been shown that in the non-relativistic, non-interacting limit, a single spin-$s$ field is indistinguishable from a set of $2s+1$ scalar fields of identical mass~\cite{Jain:2021pnk}.

In this paper we present the first three-dimensional simulations of two-field ALP models with self-interactions and inter-field interactions.  We use these simulations to study the stability of nested solitons and collisions between two solitons made from different bosonic fields.  We consider two axion-like fields with masses in the currently supported range~\cite{Marsh:2013ywa}. We also use different combinations of attractive and repulsive self-interactions and inter-field interactions. Although it is straight forward to extend these simulations to more than two-fields, this paper will focus only on two-field simulations. 

This paper is structured as follows:
In \cref{sec:eom} we present the Lagrangian, associated equations of motion and conserved energy.
In \cref{sec:code} we discuss the implementation in code.  In \cref{sec:solitons} we discuss multi-axion solitons and show their stability in said code.
In \cref{sec:collision} we present collisions between solitons in one- and two-field simulations, and show that there are qualitative differences between them.

\section{Equations of motion}
\label{sec:eom}

We assume $N$ scalar particles with Lagrangian density
\begin{equation}
\begin{split}
    \label{eq:lagrangian}
    \mathcal{L_{\text{ALP}}}
    =&
    \sum_j
    \left(
        - \frac{1}{2} g^{\mu\nu} \partial_{\mu} \phi_{j} \partial_{\nu} \phi_{j}
        - \frac{1}{2} m_{j}^2 \phi_j^2
    \right)
    \\&
    - \sum_j \sum_{k \geq j} \lambda_{jk} \phi_j^2 \phi_k^2
    \mcomma
\end{split}
\end{equation}
where the indices $j$ and $k$ run over the $N$ fields considered.
The first term describes $N$ free fields with masses $m_j$, while the second term describes interactions between them.
The symmetric matrix $\lambda_{jk}$ contains interaction constants; the diagonal elements parameterise self-interactions and the off-diagonal elements parameterize inter-field interactions.
Positive terms correspond to repulsive interactions and negative to attractive interactions.

The corresponding equations of motion for each field $\phi_j$ are
\begin{multline}
    \frac{1}{\sqrt{-g}}\partial_{\mu}\left[\sqrt{-g}g^{\mu \nu}\partial_{\nu}\phi_j\right]
        -m_j^2\phi_j-4\lambda_{jj}\phi_{j}^3
    \\
    -2\sum_{j \ne k}\lambda_{jk}\phi_j\phi_k^2=0.
\end{multline}
In the Newtonian limit the equations of motion for the $N$ fields reduce to coupled Klein-Gordon equations with self-interacting potentials,
\begin{equation}
    \label{eq:klein_gordon}
    0
    =
    \ddot{\phi_i}
    - \nabla^2 \phi_j
    + m_j^2 \phi_j 
    + 4 \lambda_{jj} \phi_j^3
    + 2 \sum_{j \ne k} \lambda_{jk} \phi_j \phi_k^2
    \mperiod
\end{equation}
In the non-relativistic limit, each of the Klein-Gordon fields $\phi_j$ can be re-written in terms of a complex scalar field $\psi_j$ in the form
\begin{equation}
    \phi_j =
    \frac{1}{\sqrt{2m}}
    \left(
    e^{-i m_j t} \psi_j + e^{+i m_j t} \psi_j^*
    \right)
\end{equation}
and \cref{eq:klein_gordon} reduces to $N$ coupled Gross-Pitaevskii-Poisson (GPP) equations
\begin{multline}
    \label{eq:gpp}
    i \hbar \frac{\partial \psi_j}{\partial t}
    =
    - \frac{\hbar^2}{2 m_j a^2} \nabla^2 \psi_j
    + m_j \Phi \psi_j
    \\
    + \frac{\hbar^3}{2 m_j^2 c} \lambda_{jj} |\psi_j|^2 \psi_j
    + \frac{\hbar^3}{4 m_j^2 c} \sum_k \lambda_{jk} |\psi_k|^2 \psi_j
\end{multline}
\begin{equation}
    \label{eq:poisson}
    \nabla^2 \Phi
    =
    \frac{1}{a} 4 \pi G \sum_j m_j |\psi_j|^2
\end{equation}
where $\Phi$ is the gravitational potential.
The first three terms on the right hand side of \cref{eq:gpp} are those of a single self-interacting field; see~\cite{Chavanis:2016dab,Glennon:2020dxs,Glennon:2022huu} for prior analysis.
The last term describes inter-field interactions.
The fields are also coupled by \cref{eq:poisson}, the Poisson equation describing gravitational interactions between the fields.

The matter density of each is equal to the modulus squared of the corresponding field,
\begin{equation}
    \rho_j
    =
    |\psi_j|^2
    \mperiod
\end{equation}

When all the masses $m_i$ are identical and there are no interactions, there is a degeneracy.  In this case, \cref{eq:gpp,eq:poisson} can alternatively be interpreted as the non-relativistic equations of motion for an integer spin-$s$ field~\cite{Jain:2021pnk,Amin:2022pzv}.
This is the case when $N \leq 2s + 1$.
The fields $\psi_i$ correspond to the components in a polarization basis.
In the non-interacting case each polarization state is conserved separately, and so $N < 2s + 1$ fields can be used to model a subset of the polarization states, assuming that the others have negligible matter content.
The case with $N=2$, $m_1 = m_2$, and $\lambda = 0$ can be interpreted as a single complex field.

The Lagrangian density giving rise to the equations of motion \cref{eq:gpp,eq:poisson} is
\begin{multline}
    \mathcal{L_{\text{GPP}}}
    =
    - \Bigg[
        \frac{1}{2} |\nabla \Phi|^2
        + \Phi \sum_j |\psi_j|^2
        + \frac{1}{2} \sum_j |\nabla \psi_j|^2
        \\
        + \frac{i}{2} \sum_j m_j \left(\psi_j \dot{\psi^*_j} - \dot{\psi_j} \psi^*_j\right)
        + \sum_j \sum_k \frac{\lambda_{jk}}{4} |\psi_j|^2 |\psi_k|^2
    \Bigg]
\end{multline}
Note that at the level of the effective Lagrangian, there is no interference term between two different fields.  Varying this Lagrangian density with respect to $\psi^*$, $\psi$, and $\Phi$, gives \cref{eq:gpp}, the conjugate of \cref{eq:gpp}, and \cref{eq:poisson}, respectively.
From this we can derive the corresponding conserved energy, 
\begin{align}
    \label{eq:energy}
    E_{\text{total}}
    &=
    \int_{\mathbb{R}^3} \dd x^3
    \left[
        \sum_j \left(
            \frac{\partial \mathcal{L}}{\partial \dot{\psi_j}} \dot{\psi_j} 
            + \frac{\partial \mathcal{L}}{\partial \dot{\psi_j^*}} \dot{\psi_j^*} 
        \right)
        \frac{\partial \mathcal{L}}{\partial \dot{\Phi}} \dot{\Phi} 
        - \mathcal{L}
    \right]
    \\
    &=
    E_{\text{grav}} + E_{\text{KQ}} + \sum_j E_{\text{self-int}, j} + \sum_j \sum_{j>k} E_{\text{int}, j, k}
\end{align}
where the gravitational potential energy is defined in the usual way,
\begin{equation}
    \label{eq:e_grav}
    E_{\text{grav}}
    =
    \frac{1}{2} \Phi \sum_j |\psi_j|^2
    \mperiod
\end{equation}
The sum of the kinetic and ``quantum''\footnote{Note that the ``quantum'' energy does not in fact have a quantum origin~\cite{Niemeyer:2019aqm}.} energy is
\begin{equation}
    \label{eq:e_kq}
    E_{\text{KQ}}
    =
    - \frac{1}{2} \sum_j \psi_j^* (\nabla^2 \psi_j)
    \mperiod
\end{equation}
The energy due to self-interactions in field $j$ is
\begin{equation}
    E_{\text{self-int}, j}
    =
    \lambda_{jj} |\psi_j|^4
\end{equation}
and the energy due to interactions between species $j$ and $k$ is
\begin{equation}
    E_{\text{int}, j, k}
    =
    \lambda_{jk} |\psi_k|^2 |\psi_j|^2
    \mperiod
\end{equation}

\section{Implementation}
\label{sec:code}

We use a modified version of \texttt{UltraDark.jl} to simulate the dynamics of multi-field ALPs~\cite{ultradark}.\footnote{\url{https://github.com/musoke/UltraDark.jl}}
\texttt{UltraDark.jl} is a pseudo-spectral solver of the GPP equations, previously used to simulate the dynamics of self-interacting fuzzy dark matter~\cite{Glennon:2022huu} and vortices in scalar dark matter~\cite{Glennon:2023oqa}.
We have extended it to simulate multiple fields and their self-interactions.

It is convenient to rewrite \cref{eq:gpp,eq:poisson} in code units, as in refs.~\cite{Edwards:2018ccc,Glennon:2020dxs,Glennon:2022huu}.
As defined elsewhere, these units depend on the mass of the (single) field.
We adapt them to use with multiple fields by writing all masses with reference to a mass $m_0$; typically $m_0 = \mathcal{O}(m_1)$ where $m_1$ is first field's particle mass.
These units are
\begin{equation}
    \mathcal{L}
    =
    {\left(\frac{8\pi \hbar^2}{3m_0^2 H_0^2 \Omega_{m0}}\right)}^{\frac{1}{4}}
    \approx
    121 {\left(\frac{10^{-23} \mathrm{eV}}{m_0}\right)}^{\frac{1}{2}} \;\mathrm{kpc},
\end{equation}
\begin{equation}
    \mathcal{T}
    =
    {\left(\frac{8\pi}{3H_0^2\Omega_{m0}}\right)}^{\frac{1}{2}}
    \approx 75.5 \;\mathrm{Gyr},
\end{equation}
and
\begin{multline}
  	  \mathcal{M}
      =
      \frac{1}{G} {\left(\frac{8\pi}{3H_0^2\Omega_{m0}}\right)}^{-\frac{1}{4}} {\left(\frac{\hbar}{m_0}\right)}^{\frac{3}{2}} 
      \\
      \approx 7\times10^7 {\left(\frac{10^{-23} \mathrm{eV}}{m_0}\right)}^{\frac{3}{2}}M_\odot.
\end{multline}

Then equations of motion are
\begin{multline}
    \label{eq:schrodinger_code}
    i \frac{\partial \psi_j}{\partial t}
    =
    - \frac{1}{2} \frac{1}{a^2} \frac{m_0}{m_j} \nabla^2 \psi_j
    + \frac{m_j}{m_0} \Phi \psi_j
    \\
    + {\left(\frac{m_0}{m_j}\right)}^2 \Lambda_{jj} |\psi_j|^2 \psi_j
    + \frac{1}{2} {\left(\frac{m_0}{m_j}\right)}^2 \sum_k \Lambda_{jk} |\psi_k|^2 \psi_j
\end{multline}
\begin{equation}
    \label{eq:poisson_code}
    \nabla^2 \Phi
    =
    \frac{1}{a} 4 \pi \sum_j \frac{m_j}{m_0} |\psi_j|^2
\end{equation}
where the interaction coefficients are written as
\begin{equation}
    \Lambda_{jk}
    =
    \frac{\hbar^2}{2 m_0^3 G \mathcal{T} c}
    \lambda_{jk}
    \mperiod
\end{equation}

In the present work we are concerned with the particular case of $N = 2$ fields, but extending the code to $N \geq 3$ fields is straightforward.
The memory requirements are roughly linear in the number of fields.
The computational complexity of each time step is roughly $\mathcal{O}(N^2)$ with inter-field interactions, $\mathcal{O}(N)$ without.

The primary constraint on extending to more fields is the range of length scales that must be resolved.
Each field $\psi_j$ has characteristic length scales roughly proportional to $1/m_j$.
Resolving these simultaneously can become challenging when there is a large spread of particle masses.  The resolution must be high enough to resolve details on the smallest length scales and the box must be large enough to accommodate the larger length scales.  The combination of a large simulation box with a fine resolution makes for very high computational costs.

\section{Multifield solitons}
\label{sec:solitons}

\begin{figure}[b]
    \centering
    \includegraphics[width=.46\textwidth,trim={0 2cm 0 0},clip]{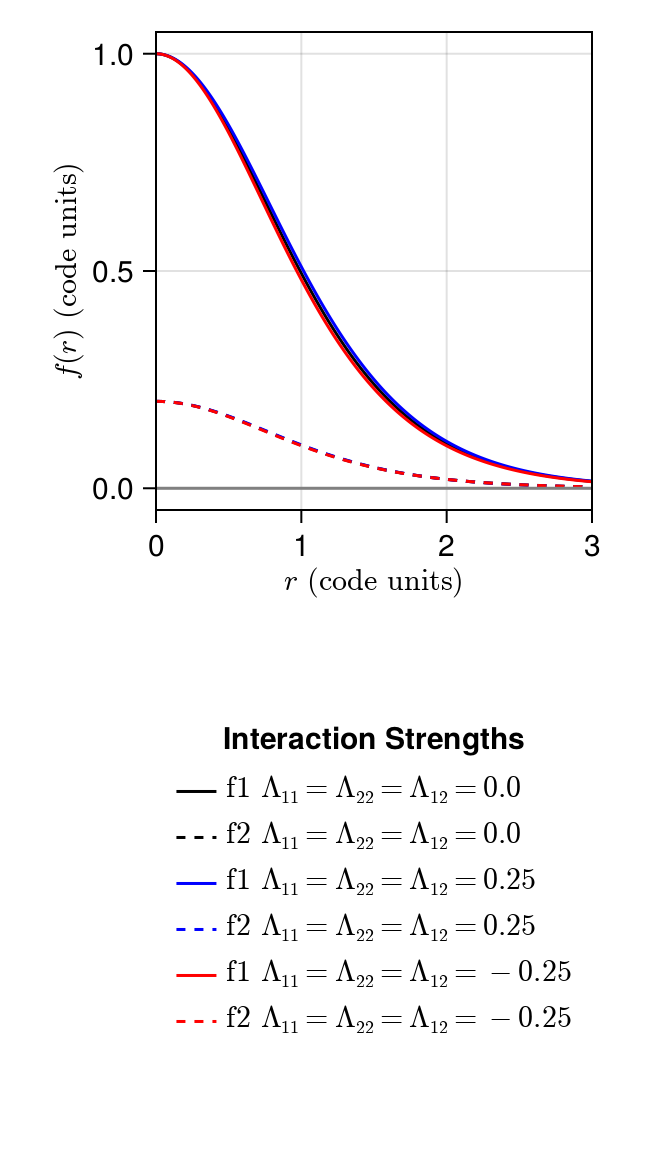}
    \caption{%
        Soliton profiles with $m_1 = m_2 = m_0$.  The solid lines represent the profile of the first solitonic field and the dashed line represents the second.
        One can see slight variations in in the initial profiles for different values of self-interactions and inter-field interactions.  The profiles are wider when there are repulsive self-interactions and inter-field interactions, and narrower when the self-interactions and inter-field interactions are attractive.
    }
    \label{m1profiles}
\end{figure}

The Gross-Pitaevskii-Poisson equations have stationary solutions called solitons.
These have been studied in great detail in both boson stars and single field FDM~\cite{Guth:2014hsa, Marsh:2015xka, Zagorac:2021qxq}.
They condense out of incoherent initial conditions~\cite{Kirkpatrick:2020fwd,Kirkpatrick:2021wwz}.  
Simulations indicate that solitons inhabit the centers of FDM halos~\cite{Du2018,Levkov2018}.

\begin{figure}
    \centering
    \includegraphics[width=.46\textwidth,trim={0 1cm 0 0},clip]{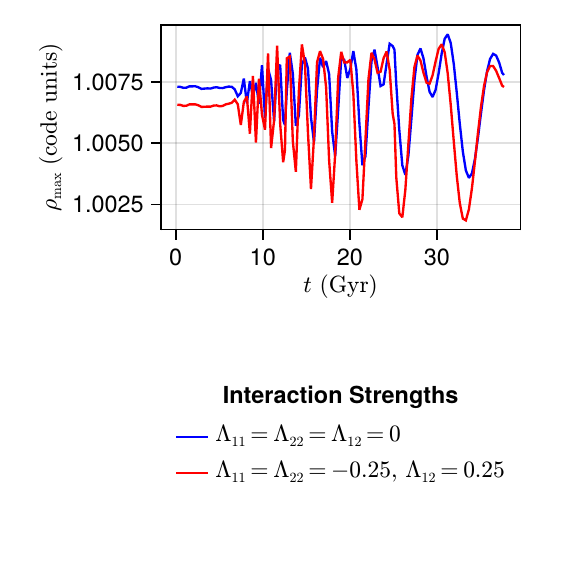}
    \caption{%
        Evolution of the maximum density of solitons over $35 \;\mathrm{Gyr}$.
        The blue curve has no self- or inter-species interactions.
        The red curve has attractive self-interactions ($\Lambda_{11} = \Lambda_{22} = -0.25$) and repulsive inter-species interactions
        In each case, the particle masses are $m_1 = 2m_2 = m_0$.  The variation in the maximum density is about 0.5\% over the duration of the simulation.
    }
    \label{DensityChange}
\end{figure}

\begin{figure}
    \centering
    \includegraphics[width=.46\textwidth,trim={0 1cm 0 0},clip]{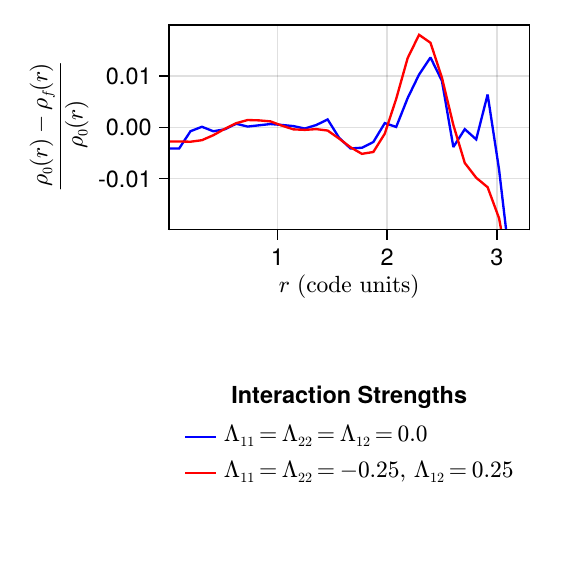}
    \caption{%
        The relative difference between the initial density profile and the density profile after $35 \;\mathrm{Gyr}$, in the same simulations as \cref{DensityChange}.  We only see percent-level deviations from the initial profiles through the central parts of the system in both the non-interacting and interacting cases.  There are larger deviations in the outer regions due to boundary effects. However, since the density is much lower in the outer regions compared to the central region, the stability of the soliton is not compromised.
    }
    \label{ProfileDiff}
\end{figure}

\begin{figure*}
    \centering
    \includegraphics[width=.95\textwidth]{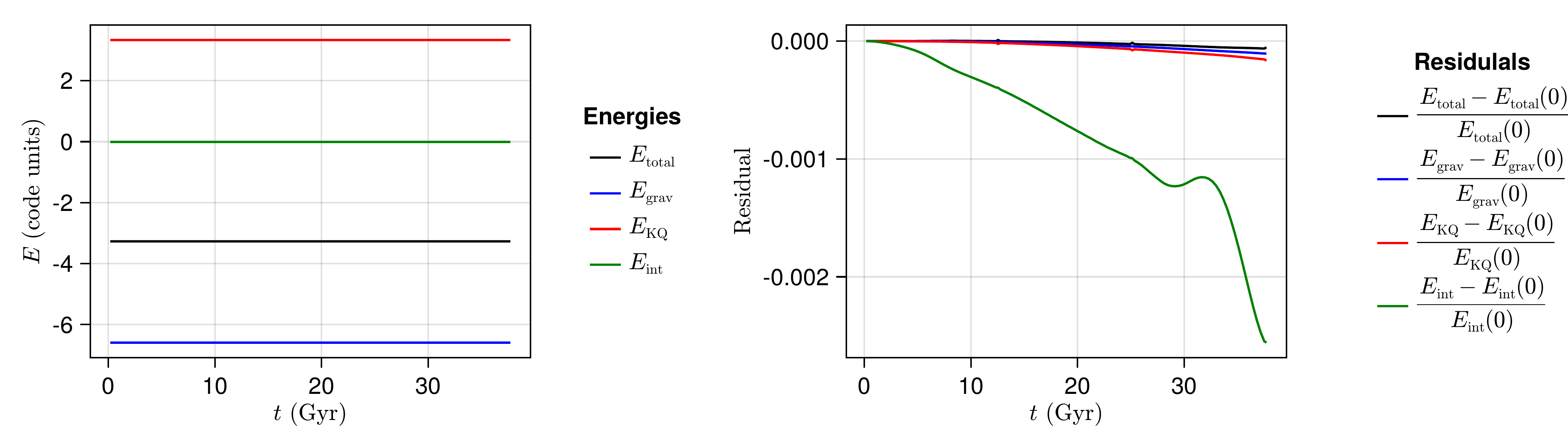}
    \caption{%
        The energy components (left) and energy residuals (right) of a sample simulation.  In these simulations, $m_1 = 2m_2 = m_0$, $\Lambda_{11} = \Lambda_{22} = -0.25$, and $\Lambda_{12} = \Lambda_{21} = 0.25$. The duration of the simulation is $35 \;\mathrm{Gyr}$.  Here, $E_{\text{int}}$ is the energy components of the self-interactions and inter-field interactions combined and $E_{\text{KQ}}$ is the sum of kinetic and ``quantum'' energies.  There is no noticeable change in the energy components over the course of the simulation.
    }
    \label{SampleEnergy}
\end{figure*}

In this paper, we use the term soliton to refer to the localized, spherically symmetric, static Bose-Einstein condensates of ALPs, possibly comprised of multiple species.  In most ULA dark matter models, solitons comprise the centers of dark matter halos which are surrounded by an incoherent outer region which is well described as a Navarro-Frenk-White (NFW;~\cite{Navarro:1996gj}) profile~\cite{Schive:2014hza, Marsh:2015xka}.  Dark matter solitons in ULA models have been studied in detail~\cite{Ruffini1969, Marsh:2015wka, Hui:2016ltb}.

We use solitons as initial conditions in our simulations.
To find these stable solitonic solutions with self-interactions and inter-field interactions, we follow a procedure similar to that used for non-interacting single species solitons in~\cite{Edwards:2018ccc,Guo:2020}.
Other procedures exist, see for example ref.~\cite{Guo:2020tla}.
We assume two scalar fields $\psi_1$ and $\psi_2$ and impose spherical symmetry of $\psi_i$ and time independence of $|\psi_i|$,
\begin{gather}
    \psi_1 \rightarrow e^{i \beta_1 t} f_1(r)
    \\
    \psi_2 \rightarrow e^{i \beta_2 t} f_2(r)
    \\
    \Phi \rightarrow \varphi(r).
\end{gather}
Combining these ansatze with \cref{eq:gpp,eq:poisson}, we find that the density profiles $f_j(r)$ and gravitational potential must be solutions to the differential equations
\begin{multline}
    f_1''(r) = -\frac{2}{r}f'_1(r) + 2\left(\frac{m_1}{m_0}\right)\tilde{\varphi}_1(r) f_1(r) 
    \\
    + 2\left(\frac{m_0}{m_1}\right) \Lambda_{11}f^3_1(r) + \left(\frac{m_0}{m_1}\right)\Lambda_{12} f^2_2(r) f_1(r)
\end{multline}
\begin{multline}
    f_2''(r) = -\frac{2}{r}f'_2(r) + 2\left(\frac{m_2}{m_0}\right) \tilde{\varphi}_2(r) f_2(r) 
    \\
    + 2\left(\frac{m_0}{m_2}\right) \Lambda_{22}f^3_2(r) + \left(\frac{m_0}{m_2}\right)\Lambda_{12} f^2_1(r) f_2(r)
\end{multline}
\begin{equation}
    \tilde{\varphi}''(r) = 4\pi\left(\left(\frac{m_1}{m_0}\right)f^2_1(r)+\left(\frac{m_2}{m_0}\right)f^2_2(r)\right) - \frac{2}{r}\tilde{\varphi}'(r)
\end{equation}
where $\tilde{\varphi}_i(r) = \left(\frac{m_i}{m_0}\right)\left(\varphi(r) + \beta_i\right)$ are rescaled gravitational potentials.
Note that $\tilde{\varphi}''(r) = \left(\frac{m_0}{m_1}\right)\tilde{\varphi}_1''(r) = \left(\frac{m_0}{m_2}\right)\tilde{\varphi}_2''(r)$ and $\tilde{\varphi}'(r) = \left(\frac{m_0}{m_1}\right)\tilde{\varphi}_1'(r) = \left(\frac{m_0}{m_2}\right)\tilde{\varphi}_2'(r)$ so the last equation can be written in terms of derivatives of either $\tilde{\varphi}_1$ or $\tilde{\varphi}_2$.

Not all solutions to these equations are solitons.
Most of them have $\lim_{r \to \infty} f(r) = \pm \infty$; these solutions have infinite mass.
In order to find physical solutions, one must choose sensible initial conditions $f_j(0)$, $f_j'(0)$, $\varphi(r)$, $\varphi'(0)$.
In every case we consider, $f_j'(0) = \varphi'(0) = 0$.
This is because we assume the soliton is in its ground state, with a local maximum at $r=0$.  The central densities $f_j(0)$ are set by the desired soliton mass.
To find suitable values for $\varphi_1$ and $\varphi_2$, we search for those for which 
\begin{equation}
    \lim_{r \to \infty} f(r) = 0
\end{equation}
and $f(r)$ has no nodes.
Our algorithm uses a modified shooting method to find such solutions for $0.1 \lesssim m_1/m_2 \lesssim 10$ and $|\Lambda_{ij}| \lesssim 1$.

In Figure~\ref{m1profiles}, we show profiles for solitons comprised of two fields, when the particle mass of each field is the same but they have differing self-interactions and inter-field interactions.
Since we generated the initial profiles assuming the central density is the same in each profile, the cores of the solitons look similar.
The differences in the profiles exist mostly in the outer regions of the solitons.
As expected, introducing attractive interactions with $\Lambda_{ij} < 0$ causes the equilibrium soliton to contract; repulsive interactions with $\Lambda_{ij} > 0$  cause it to expand.
While these differences appear subtle, if we were to assume an equilibrium profile with no self-interactions, we would see significant oscillations in supposedly static solutions (approximately $5\%$ of the peak density) when there are attractive or repulsive self-interactions.

We used the resulting soliton solutions to evaluate the correctness of our multi-field modifications to \texttt{UltraDark.jl}.
We initialized profiles with a variety of particle mass ratios, self-interactions, and inter-field interactions and evolved them forward to see if they were in fact equilibrium solutions.
For all the initial profiles, we assume that the central densities are $f_1(0) = 1.0$ and $f_2(0) = 0.2$; this ratio is chosen such that each field provides a significant but distinct contribution.

\Cref{DensityChange,ProfileDiff} shows representative tests that solitons are equilibria.
There are two cases shown: one with no self-interactions ($\Lambda_{ij} = 0$) and one with attractive self-interactions ($\Lambda_{11} = \Lambda_{22} = -0.25$) and repulsive inter-species interactions ($\Lambda_{12} = 0.25$).
Both have  $m_1=2m_2=m_0$.
Figure~\ref{DensityChange}, shows how the central density of the overlapping solitons changes over time.
This figure shows that the oscillations in the profile are small, meaning the profile we initialize is very close to equilibrium and the code preserves it.
Figure~\ref{ProfileDiff} shows the fractional change in the density profiles after the soliton has evolved forward for $35 \;\mathrm{Gyr}$.
Note that the amplitude in Figure~\ref{ProfileDiff} depends on the time when you measure the density profile, but $35 \;\mathrm{Gyr}$ is representative.
This figure shows that there are only slight deviations from the initial density profile even after a significant amount of time has passed.
The largest relative deviations are in the low-density exterior of the soliton.
This is largely due to the periodic boundary conditions: the soliton feels gravitational forces due to neighboring boxes.

In Figure~\ref{SampleEnergy}, we show how the energy components evolve over time for a set-up where $m_1 = 2m_2 = m_0$, there are attractive self-interactions with $\Lambda_{11} = \Lambda_{22} = -0.25$, and repulsive inter-field interactions with $\Lambda_{12} = \Lambda_{21} = 0.25$.  The energy components are each conserved over the duration of $35 \;\mathrm{Gyr}$, further indicating that the soliton is near equilibrium.  We see the same stability in simulations with different particle mass ratios, self-interactions and inter-field interactions.

\section{Collisions between solitons}
\label{sec:collision}

Collisions between solitons in single field FDM are well studied~\cite{Schive:2014hza,Schwabe:2016rze,Edwards:2018ccc,Hertzberg:2020dbk,Glennon:2020dxs,Jain:2021pnk}.
Binary collisions have been used to study the basic dynamics of FDM fields and demonstrate effects such as self-interactions.
Mergers of larger numbers of solitons have been used as a proxy for halo formation through hierarchical mergers~\cite{Schive:2014hza,Schwabe:2016rze,Zagorac:2021qxq}.

\begin{figure*}
    \centering
    \includegraphics{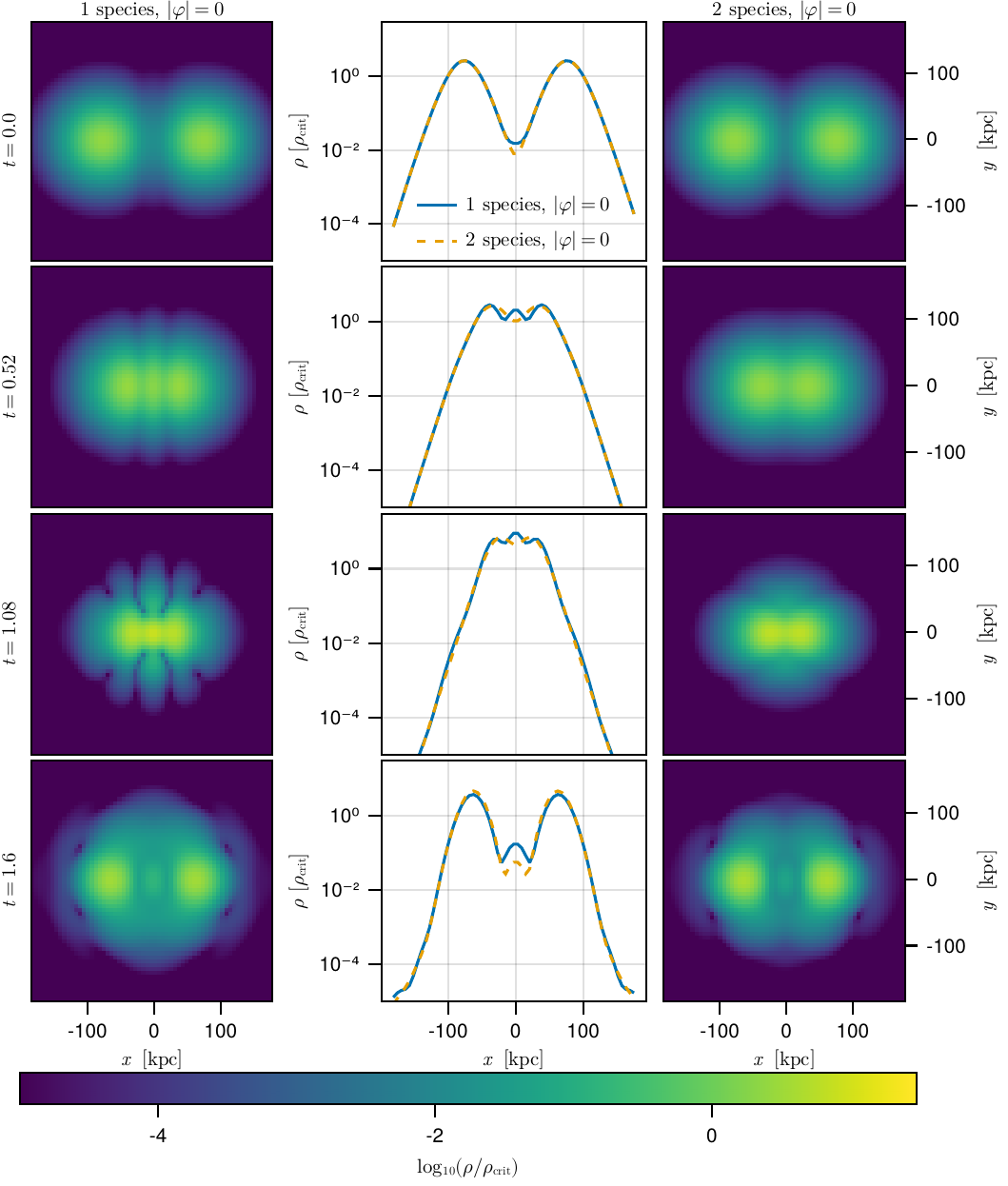}
    \caption{%
        Snapshots of a collision between a pair of unbound solitons with identical phases.
        Left: Both solitons are in one field.
        Right: The solitons are in distinct fields with $m_1=m_2$.
        Middle: Density along the axis of the collision for the one species collision in solid blue and two species collision in dashed orange.
        One can see that in the case with one species, the solitons interfere when they overlap.
        In the case with solitons comprised of different species, they do not interfere as they pass through each other.
        This is a qualitative difference between 1- and 2-species ALPs.
        See \url{https://www.youtube.com/watch?v=IENq5imeIzE} or \url{https://doi.org/10.5281/zenodo.7675774} for an animation.
    }%
    \label{fig:unbound}
\end{figure*}

\begin{figure*}
    \centering
    \includegraphics{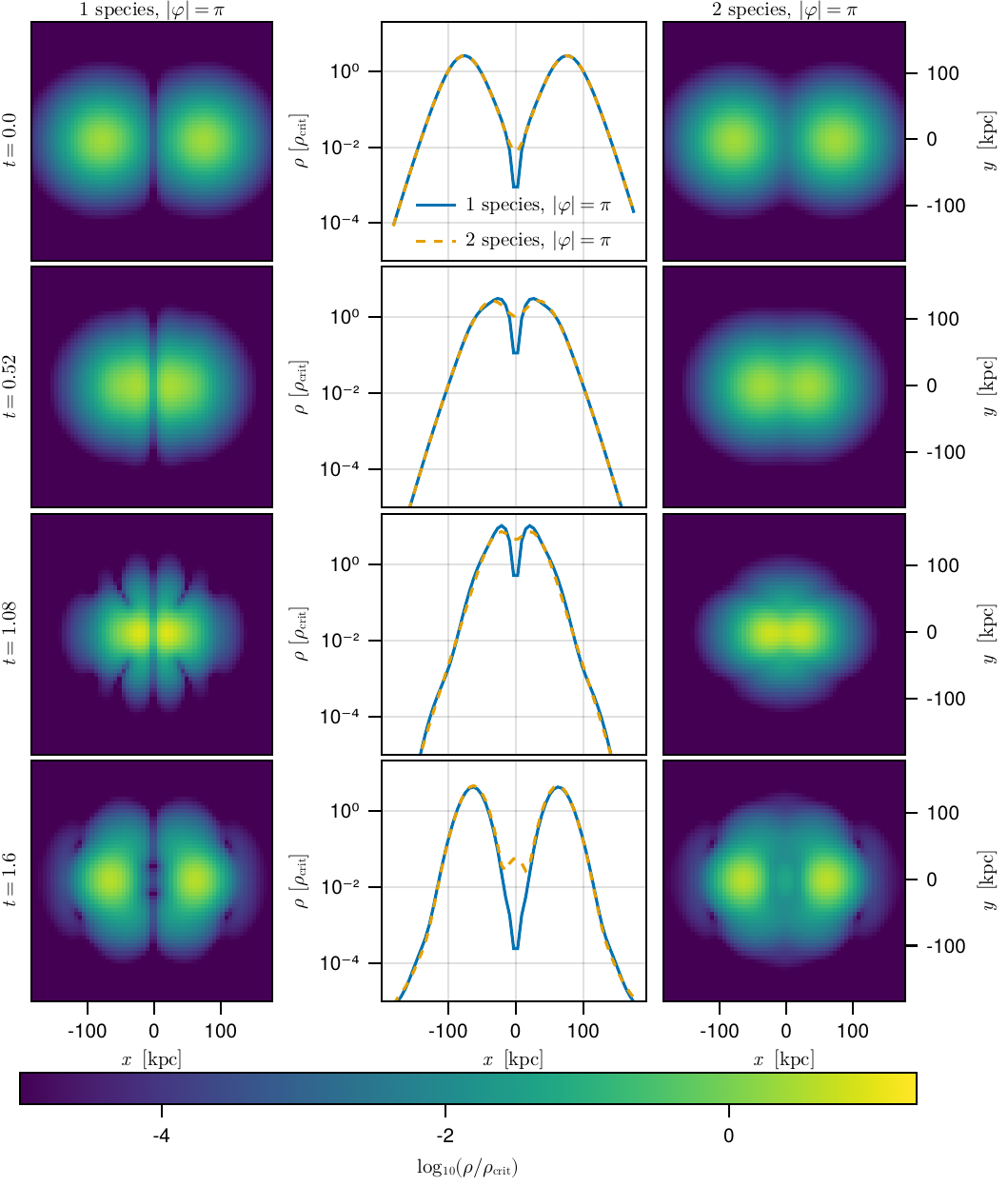}
    \caption{%
        Snapshots of a collision between a pair of unbound solitons a phase difference of $\pi$.
        Left: Both solitons are in one field.
        Right: The solitons are in distinct fields with $m_1=m_2$.
        Middle: Density along the axis of the collision for the one (two) species collision in solid blue (dashed orange).
        One can see that in the case with one species, interference means that the density at the $x=0$ plane between the solitons is always a local minimum.
        In the case with solitons comprised of different fields they do not interfere and there is instead a maximum at the midpoint when they pass through each other.
        See \url{https://www.youtube.com/watch?v=vJhFGEdrLzw} or \url{https://doi.org/10.5281/zenodo.7675774} for an animation.
    }%
    \label{fig:phase-diff}
\end{figure*}

\begin{figure*}
    \centering
    \includegraphics{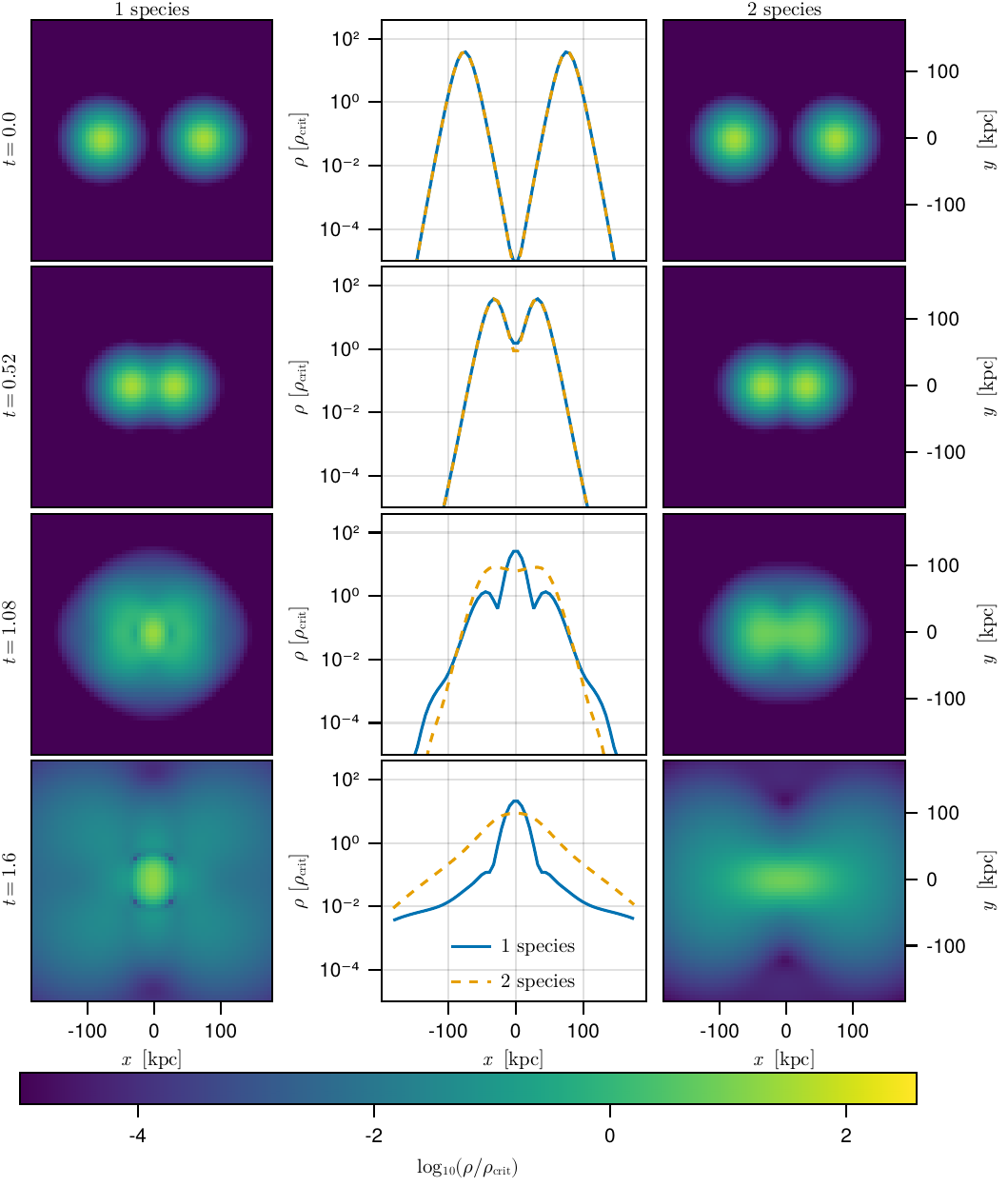}
    \caption{%
        Snapshots of collisions between gravitationally bound solitons in 1- and 2-species scenarios.
        Left: Density slices in single field collision.
        Middle: Density along line through axis of collision for one (solid blue) and two (dashed orange) fields.
        Right: Density slices in two field collision.
        The one-field scenario shows interference fringes and a more sharply peaked core in the end state.
        See \url{https://www.youtube.com/watch?v=_VaDPWeVVdo} or \url{https://doi.org/10.5281/zenodo.7675774} for an animation.
    }
    \label{fig:merger_1_2_snapshots}
\end{figure*}

\begin{figure*}
    \centering
    \includegraphics{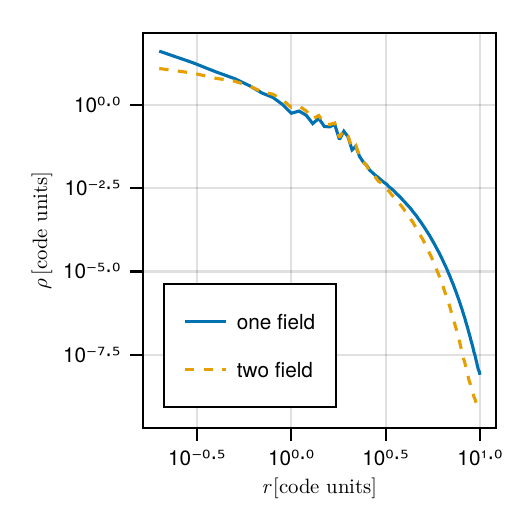}
    \includegraphics{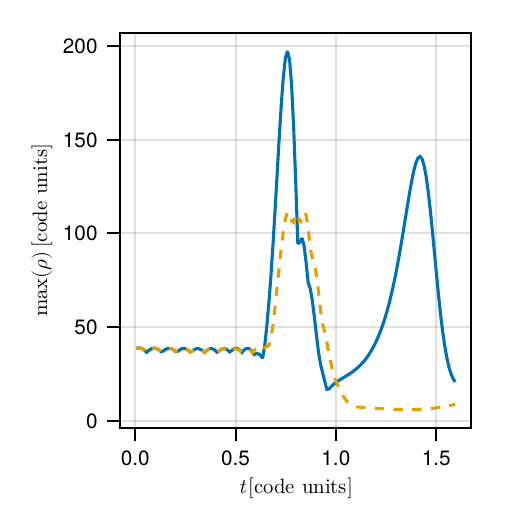}
    \includegraphics{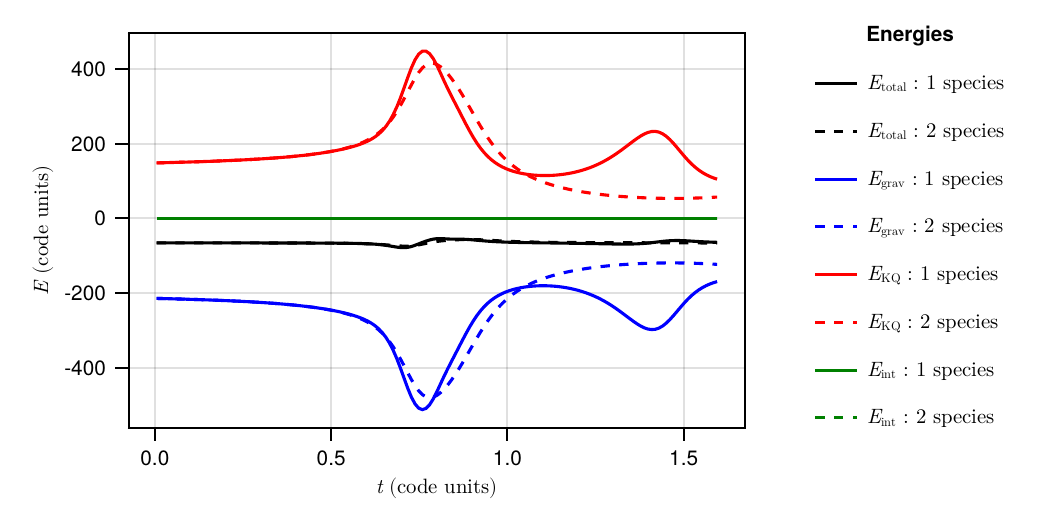}
    \caption{%
        Results of colliding two equal mass solitons from the same and different species, as in \cref{fig:merger_1_2_snapshots}.
        Top left:
        Spherically averaged density profiles of the 1-species (solid blue) and 2-species (dashed orange) cases, time averaged over the end of the simulation.
        The single species profile is more sharply peaked.
        Top right:
        Maximum density as a function of time.
        Although both scenarios have significant oscillations in the density, the two-species scenario has a smaller amplitude.
        Bottom:
        Time series of the energy for the same pair of simulations.
        The single-species merger has more kinetic energy after the collision.
    }
    \label{fig:merger_1_2_profiles}
\end{figure*}

\begin{figure*}
    \centering
    \includegraphics{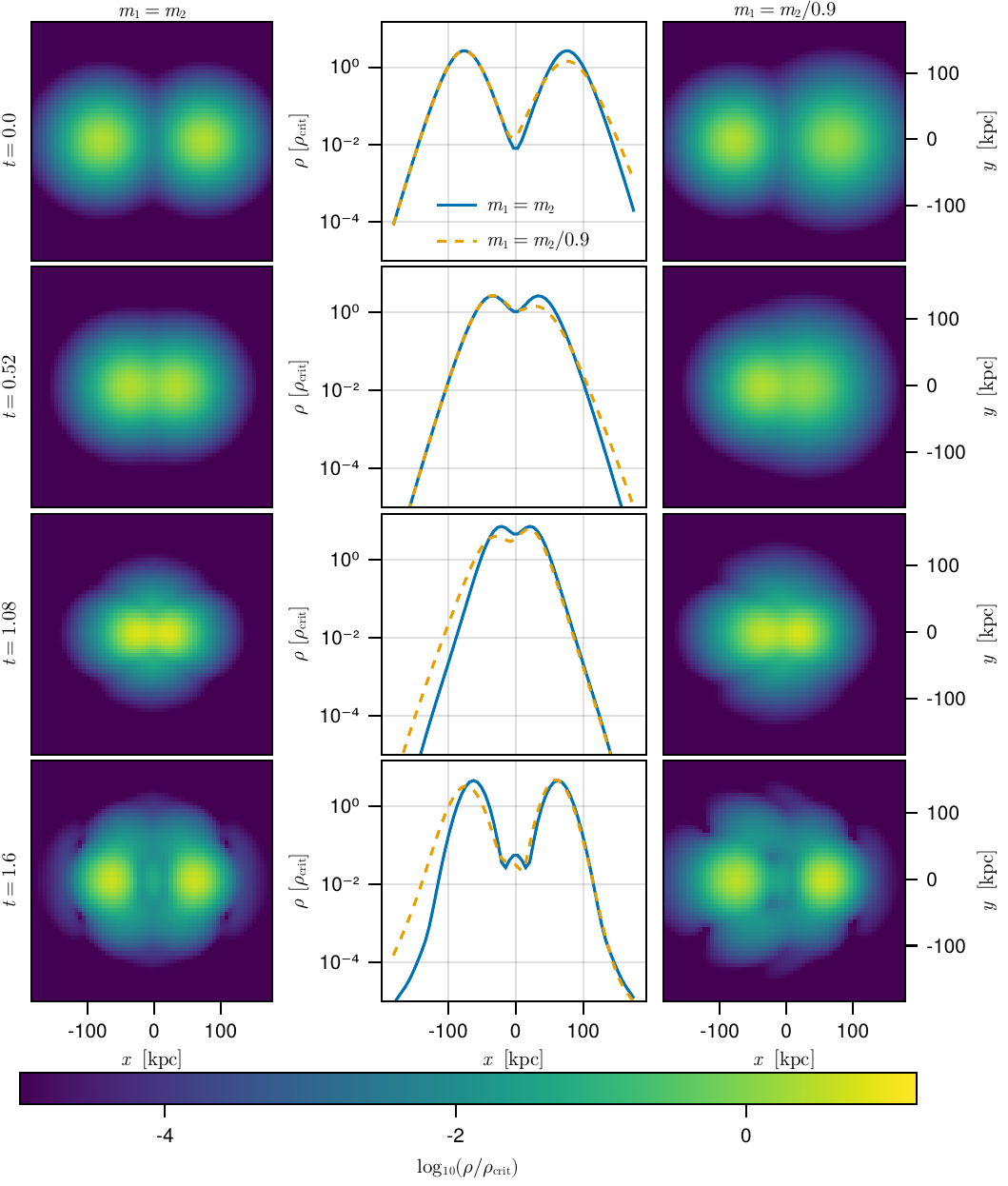}
    \caption{%
        Snapshots of collisions between pairs of unbound solitons.
        Left: The solitons have distinct constituent species with $m_1 = m_2$.
        Right: The solitons have distinct constituent species.  The soliton initially on the left (right) is composed of particles with mass $m_1$ ($m_2 = 0.9 \times m_1$).
        The lighter particle has a longer characteristic wavelength, so an equal mass soliton is more extended.
        Middle: Density along the axis of the collision, corresponding to the left and right columns in solid blue and dashed orange, respectively.
        One can see that in the case with $m_1 \ne m_2$, there is greater asymmetry in the end state of the collision.
        See \url{https://www.youtube.com/watch?v=9yya57eDV4U} or \url{https://doi.org/10.5281/zenodo.7675774} for an animation.
    }%
    \label{fig:unequal_mass}
\end{figure*}

\begin{figure*}
    \centering
    \includegraphics{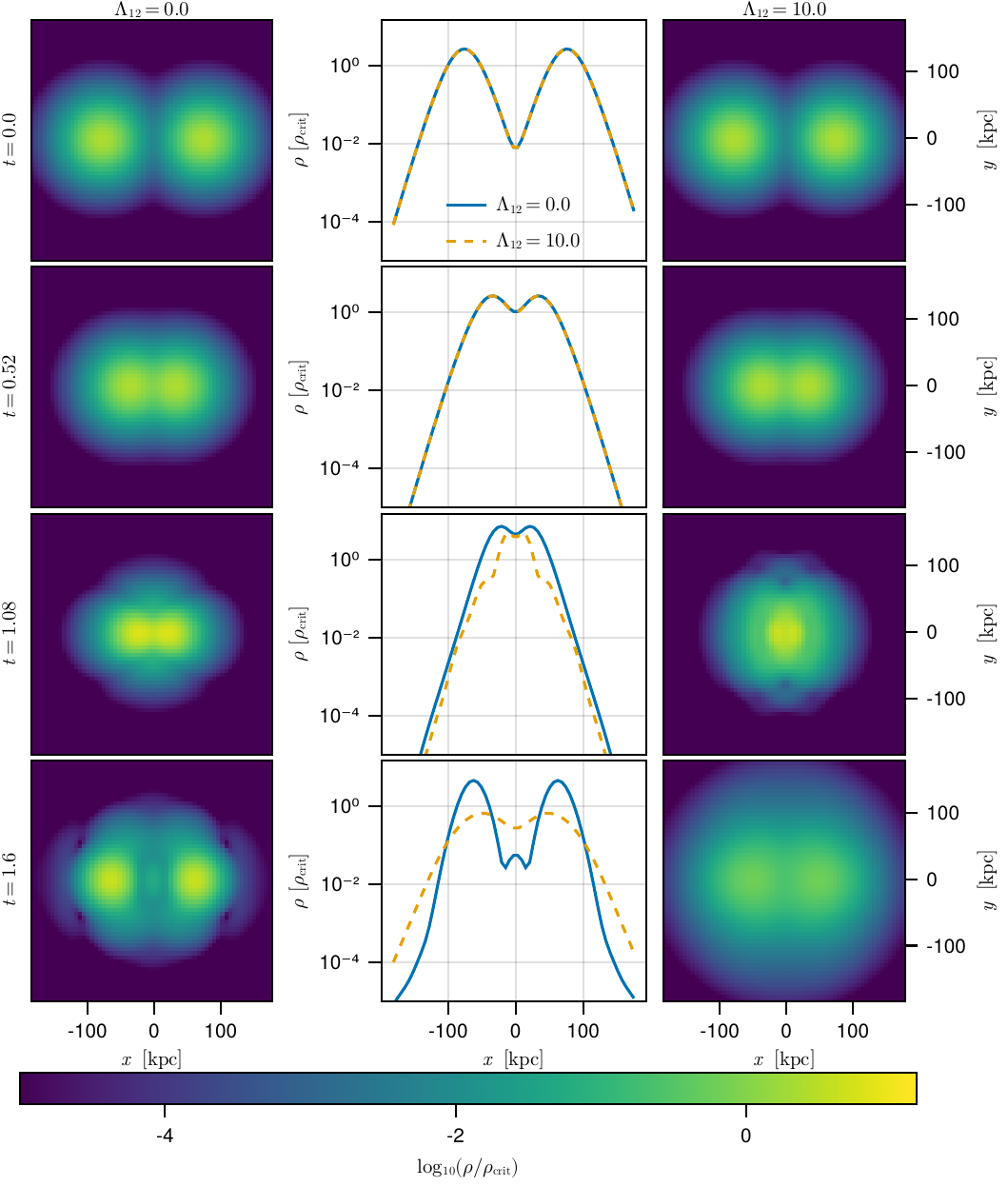}
    \caption{%
        Snapshots of collisions between pairs of unbound solitons.
        Left: The solitons have different constituent fields, with $m_1 = m_2$ and $\Lambda=0$.
        Right: The solitons are in distinct fields with $m_1 = m_2$.
        There is a repulsive inter-species interaction with $\Lambda_{12} = 10$.
        Middle: Density along the axis of the collision, corresponding to the left and right columns in solid blue and dashed orange, respectively.
        The solitons with $\Lambda_{12} = 10$ have strongly repulsive inter-field interactions.
        Instead of passing through each other with small perturbations, they are split into two components and their peak density is greatly suppressed after the collision.
        See \url{https://www.youtube.com/watch?v=AyQe4coQXf8} or \url{https://doi.org/10.5281/zenodo.7675774} for an animation.
    }%
    \label{fig:repulsive}
\end{figure*}

\begin{figure*}
    \centering
    \includegraphics{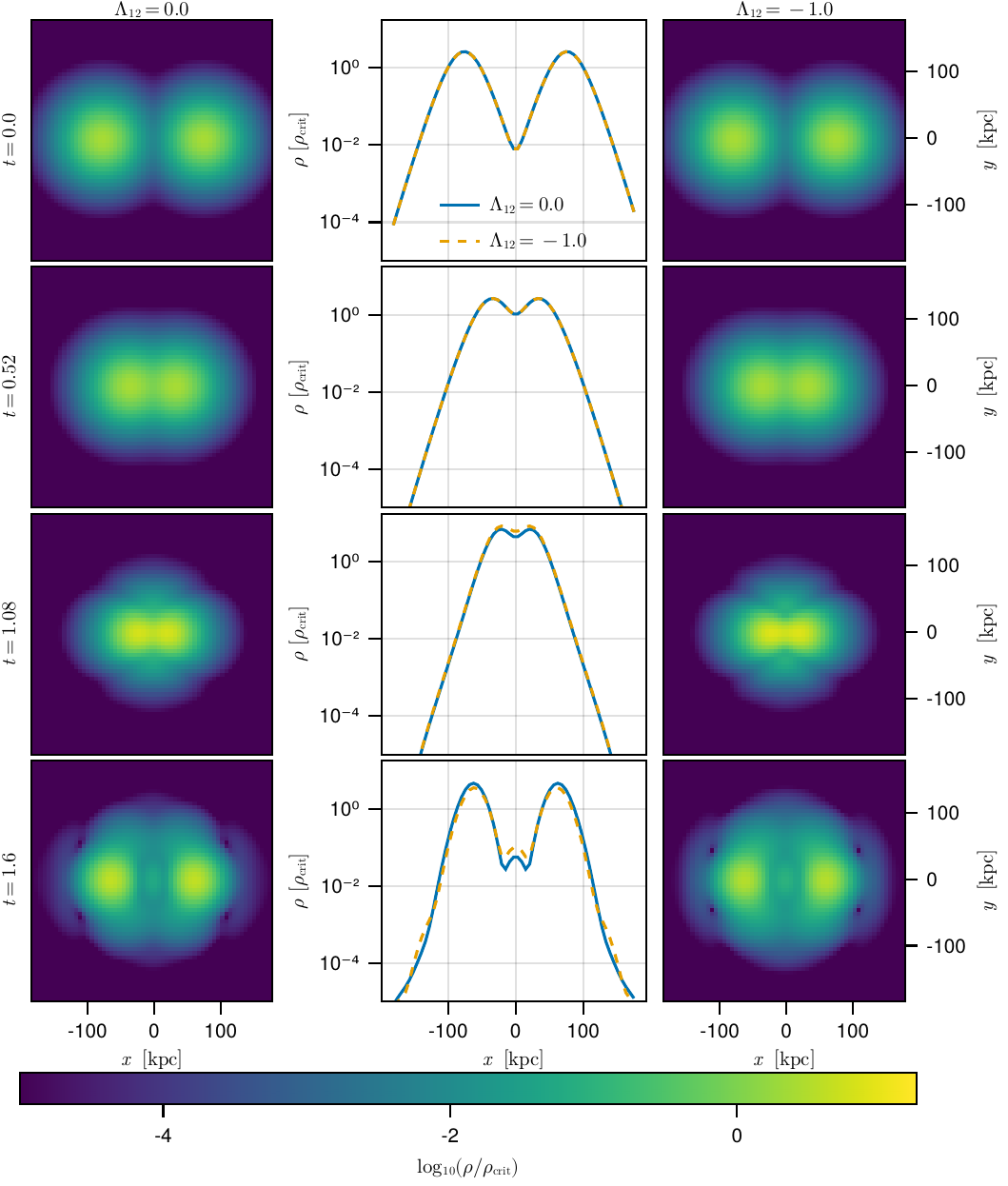}
    \caption{%
        Snapshots of collisions between pairs of unbound solitons.
        Left: The solitons have different constituent fields, with $m_1 = m_2$ and $\Lambda=0$.
        Right: The solitons are in distinct fields with $m_1 = m_2$.
        There is a attractive inter-species interaction with $\Lambda_{12} = -1$.
        Middle: Density along the axis of the collision, corresponding to the left and right columns in solid blue and dashed orange, respectively.
        The solitons with $\Lambda_{12} = -1$ have strongly attractive inter-field interactions.
        The peak between the solitons in the end state is enhanced by attractive interactions.
        See \url{https://www.youtube.com/watch?v=W5KNJb94wDI} or \url{https://doi.org/10.5281/zenodo.7675774} for an animation.
    }%
    \label{fig:attractive}
\end{figure*}

\begin{table}[]
    \centering
    \begin{tabular}{c|c|c|c|c|c}
         $N$ & $m_1/m_2$ & $M_s$ & $\Lambda_{12}$ & $|\varphi|$ & figure
         \\ \hline
         1 & 1 & $5 \mathcal{M}$ & 0 & 0 & \ref{fig:unbound}
         \\
         2 & 1 & $5 \mathcal{M}$ & 0 & 0 & \ref{fig:unbound}, \ref{fig:unequal_mass}, \ref{fig:repulsive}, \ref{fig:attractive}
         \\
         1 & 1 & $5 \mathcal{M}$ & 0 & $\pi$ & \ref{fig:phase-diff}
         \\
         2 & 1 & $5 \mathcal{M}$ & 0 & $\pi$ & \ref{fig:phase-diff}
         \\
         1 & 1 & $10 \mathcal{M}$ & 0 & 0 & \ref{fig:merger_1_2_snapshots}, \ref{fig:merger_1_2_profiles}
         \\
         2 & 1 & $10 \mathcal{M}$ & 0 & 0 & \ref{fig:merger_1_2_snapshots}, \ref{fig:merger_1_2_profiles}
         \\
         2 & 0.9 & $5 \mathcal{M}$ & 0 & 0 & \ref{fig:unequal_mass}
         \\
         2 & 1 & $5 \mathcal{M}$ & 10 & 0 & \ref{fig:repulsive}
         \\
         2 & 1 & $5 \mathcal{M}$ & -1 & 0 & \ref{fig:attractive}
    \end{tabular}
    \caption{%
        Summary of parameters in numerical experiments performed.
        When $N=1$, both solitons have the same field; when $N=2$, each initial soliton has a different field.
        The ratio of particle masses is $m_1/m_2$ .
        The initial solitons have masses $M_s$.
        The inter-species interaction strength is given by $\Lambda_{12}$ and the phase difference by $|\varphi|$.
        Also indicated is the figures in which a given scenario is plotted.
    }
    \label{tab:params}
\end{table}

We continue this tradition of using soliton dynamics to elucidate properties of FDM\@.
In an effort to simplify comparisons between plots in the following scenarios, we look at the effects of each of multi-component ALPs' properties separately: multiple species, distinct particle masses $m_i$ for each species, and inter-species interactions $\Lambda$.
In each scenario the initial conditions contain two solitons, each composed of a single species and,  unless otherwise specified, phase difference $\varphi = 0$.
The scenarios have some common parameters:
the solitons start $4 \,\mathcal{L}$ apart, with velocities $v_1 = -v_2 = 2 \mathcal{L}/\mathcal{T}$.
There are two general classes of collisions: those in which the solitons are gravitationally bound and unbound.
In the unbound scenarios, the mass of each soliton is $5 \,\mathcal{M}$.
In the bound scenarios, the mass of each soliton is $10 \,\mathcal{M}$.
These parameters were chosen to capture a wide variety of phenomena, rather than correspond to a specific physical scenario, and are summarized in \cref{tab:params}.
For clarity, each set of snapshots is cropped to the interior of the box; see the linked animations for uncropped versions.\footnote{\url{https://www.youtube.com/playlist?list=PLHrf0iQS5SY5-pjTrIWMDfelGTEvd48l0}, also archived at \url{https://doi.org/10.5281/zenodo.7675774}\cite{zenodo_supplement}}

One of the most distinctive FDM effects is interference~\cite{Schive:2014dra}, which can manifest when solitons overlap during collisions.
In the collision of same-species solitons, their wave-like nature shows up as a distinctive interference pattern in the density when they overlap~\cite{Schwabe:2016rze},
\begin{equation}
    \begin{split}
        \psi(x, t)
        =&
        A_1 \left(|x+\hat{x}|\right)e^{i(kx/2+\omega t+\varphi/2)}
        \\
        & + A_2 \left(|x-\hat{x}|\right)e^{i(-kx/2+\omega t-\varphi/2)}
        \mperiod
    \end{split}
\end{equation}
Here, $\pm \hat{x}$ are the starting positions of the solitons, ${(A_1)}^2$ and ${(A_2)}^2$ are the density profiles of the solitons, $\varphi$ is the relative phase between the solitons, and $k = mv_{||}/\hbar$ is the wave number associated with the relative velocity of the solitons. This does not happen when the two solitons are comprised of different FDM species; their densities are added rather than their wave functions.
This means that collisions between solitons consisting of a single species are necessarily different from collisions between solitons comprised of different species.
Jain \emph{et al.}~\cite{Jain:2021pnk} showed that the analogous collisions between solitons in distinct polarized states are a potential signature of higher-spin ultralight dark matter: unless the solitons are fine tuned, the density during collisions between solitons is dependent on whether they are comprised of the same field and/or polarization.

\Cref{fig:unbound} demonstrates this difference in interference effects collisions between unbound solitons with phase difference $\varphi = 0$.
In one scenario, on the left, both solitons are comprised of the $\psi_1$ field.
In the other scenario, one soliton is comprised purely of the $\psi_1$ field, and the other of the $\psi_2$ field with $m_1 = m_2$.
Both cases have no self-interactions; $\lambda = 0$.
In each case, the solitons pass through each other with some deformation, the form of which depends on whether they are comprised of the same species.
There are clear interference fringes when the solitons are comprised of the same species.
This interference does not happen if the solitons consist of distinct species.
Instead, the deformations of the solitons during the collision are entirely due to gravitational interactions.

A relative phase of $\varphi = \pi$ between solitons can cause them to bounce off one another during collisions, rather than passing through one another or merging~\cite{Edwards:2018ccc,Glennon:2020dxs}.
This can be viewed as an extreme case of the interference discussed above.
As above, different species do not interfere with each other, and so this does not occur in collisions between solitons comprised of different species.
We demonstrate this difference between single- and multi-species FDM in \cref{fig:phase-diff}.
Because the phases are exactly opposite, the density  vanishes in the $x=0$ plane separating the solitons in the single species case.
In contrast, the density in the two species case is identical to when the phases are equal.

\Cref{fig:merger_1_2_snapshots} contrasts the results of colliding bound solitons.
As above, one case has both solitons comprised of the $\psi_1$ field.
The other case has a soliton from each of the $\psi_1$ and $\psi_2$ fields with $m_1 = m_2$ and $\lambda = 0$.
The solitons merge to form a core surrounded by an NFW-like skirt and some mass is ejected.
The details of the merger depend on factors including their relative masses, velocities, and phases and self-interactions in the field.
As in the unbound scenario, the single field solitons display distinct interference fringes when they overlap.
However, there is a difference: the dark matter halos in the end state of the merger have different density profiles.
In \cref{fig:merger_1_2_profiles}, we show that the resulting density profile is less peaked in the two-field scenario.
This is consistent with the findings that the density field of multi-species FDM is smoother~\cite{Gosenca:2023yjc} and that collisions of vector dark matter solitons result in less dense cores than their scalar dark matter counterparts~\cite{Amin:2022pzv}.
The evolution of the energy components is also different: the single species takes longer to dissipate kinetic energy.

Next we examine collisions between solitons whose constituent particles have unequal mass.
For simplicity of illustration, we assume the masses are not extremely different.
\Cref{fig:unequal_mass} shows a comparison between a collision between solitons with $m_1 = m_2$, and a collision with $m_2 = m_1 \times 0.9$.
Note that although the central density of the $\psi_2$ soliton is significantly lower than that of the $\psi_1$ soliton, they contain equal total masses.
The $\psi_2$ soliton is comprised of particles with a larger de~Broglie wavelength, and so has a larger characteristic radius.
This introduces larger radial distortions in the solitons, than when $m_1 = m_2$.
This relatively small difference in particle mass is also sufficient to introduce an offset in the soliton position along the $x$-axis, relative to the equal-mass case.
A larger difference in ALP masses will enhance these differences in collision dynamics.

In \cref{fig:repulsive} we compare collisions between solitons with no interactions ($\Lambda_{ij} = 0$) and with strongly repulsive inter-species interactions $\Lambda_{12} = 10$.
Repulsive interactions suppress the density of the solitons immediately after the collision, as compared to collisions with $\Lambda_{12} = 0$.
We are not aware of a scenario where this happens in collisions of single field ALPs.
When the repulsive inter-species interaction is strong enough, there is another qualitative difference from collisions with $\Lambda = 0$: instead of passing through each other with perturbations, the solitons are split.
In \cref{fig:attractive} we compare collisions between solitons with no interactions ($\Lambda_{ij} = 0$) and with strongly attractive inter-species interactions $\Lambda_{12} = -1$.
The density peak at the midpoint between the solitons is larger than when there are no inter-species interactions.
The maximum density during the collision is also increased from $7.5 \; \mathcal{M}/\mathcal{L}^3$ to $8.9 \; \mathcal{M}/\mathcal{L}^3$.
For strong enough attractive self-interactions, an otherwise unbound system becomes bound.

\section{Discussion}

We have taken initial steps to simulate the dynamics of self- and inter-species interacting multi-species models of FDM.
These models are motivated by the ``axiverse'' conjecture, in which there are numerous axion-like particles, each with its own mass and self-interactions~\cite{Arvanitaki:2009fg}.
These are also applicable to spin-$s$ FDM models, in which each the field can be decomposed into $2s+1$ fields.
We have verified the integrity of our code by finding stable equilibrium solutions when there are multiple-fields with different masses and interaction strengths.
We have made an exploratory study of binary collisions with different combinations of soliton phases, soliton masses, particle masses and interaction strengths.

It is important to note that the collision scenarios we have discussed in this paper are highly idealized, and in the real world, collisions are more complex.
First, FDM solitons are not expected to be in zero background density regions; after initial condensation they would be surrounded by a nearly homogeneous background density and later they may have NFW-like tails.
Secondly, we have little reason to believe that multi-field ALPs would form solitons or halos comprised purely of a single field.
The multi-field nature could still show up in collisions.
For example, the phases of fields in a halo do not have to be correlated.
This could lead to one species being ejected during a merger while others are not, resulting in a halo with a deficiency or surplus of a single species.

There are a variety of future directions to be explored.
The most obvious is the generalization of these simulations from two to three and more FDM species.
Simulations of three FDM species would allow more direct comparisons with spin-$1$ fields~\cite{Jain:2021pnk,Amin:2022pzv,Jain:2022agt}.
Other directions for immediate future work would further explore the dynamics of two fields.

Structure formation and the initial condensation of solitons is different when dark matter is comprised of multiple species of ultralight particle.
In ref.~\cite{Gosenca:2023yjc}, the authors found that as multiple axion fields evolved without self-interactions or inter-field interactions, the fields showed little correlation.
It would be interesting to see how this changes when there are non-zero inter-field interactions.
Likewise, structure formation and soliton condensation in the presence of multiple fields has the potential to be quite different from a single field.
Repulsive inter-field interactions raise the possibility that the fields separate during the initial condensation process, leading to an inhomogeneous distribution of dark matter species.
This effect would leave an imprint in large scale structure, implying constraints on repulsive inter-field interactions.

There are numerous unexplored avenues for multi-field solitons.
We will generalise our algorithm for generating equilibrium profiles, to allow for larger values of particle mass ratios $m_1/m_2$, as well as allow for greater self-interaction and inter-field interaction strengths.
There are also more complex soliton interactions and collisions to be explored.
In this work, we have not explored collisions between solitons that are initially comprised of multiple fields.
We can also explore collisions with non-zero impact parameters or collisions with more than two solitons.
We hope that future work on this topic will lead to insight on the validity of the axiverse hypothesis.

\acknowledgments{}

We would like thank
Arka Banerjee,
David Cyncynates,
Neal Dalal,
Richard Easther,
Ethan Nadler,
Mark Neyrinck,
Olivier Simon,
L.~C.~R.~Wijewardhana, and Luna Zagorac
for helpful discussions.
We would also like to thank the custodial and administrative staff at the University of New Hampshire including Katie Makem-Boucher and Michelle Mancini.

Computations were performed on Marvin, a Cray CS500 supercomputer at UNH supported by the NSF MRI program under grant AGS-1919310.
This work was performed in part at Aspen Center for Physics, which is supported by National Science Foundation under Grant No. PHY-1607611. NG's participation was supported in part by the National Science Foundation under Grant No. 1929080.

\bibliographystyle{apsrev4-1}
\bibliography{autobib,manualbib}

\end{document}